# Nanostructured electrodes for thermionic and thermotunnel devices

Avto N. Tavkhelidze

*Tbilisi State University, Chavchavadze ave. 13, Tbilisi 0179, Georgia*
*E-mail: avtotav@gmail.com*

Recently, new quantum features have been studied in the area of ridged quantum wells (RQW). Periodic ridges on the surface of the quantum well layer impose additional boundary conditions on the electron wave function and reduce the quantum state density. Electrons, rejected from forbidden quantum states, have to occupy the states with higher energy. As a result, Fermi energy in RQW increases and work function (WF) decreases. We investigate low WF electrode, composed from a metal RQW layer and a base substrate. The substrate material was selected so that electrons were confined to the RQW. The WF value depends on ridge geometry and electron confinement. We calculate WF in the metal RQW films grown both on a semiconductor and metal substrates. In the case of semiconductor substrate, wide band gap materials are preferable as they allow more reduction in RQW work function. In the case of metal substrate, low Fermi energy materials are preferable. For most material pairs, the WF was reduced dramatically. Such structures, can serve as electrodes for room temperature thermionic and thermotunnel energy converters and coolers.

## I. INRODUCTION

Low work function (WF) electrodes[1] are essential for cold emission and room temperature operation of thermionic [2-6] and thermotunnel [7-13] energy converters and coolers. Such electrodes require materials with work function $e\phi = 0.2 - 0.4$ eV (here, $e$ is electron charge and $\phi$ potential). For most metals $e\phi > 4$ eV, and only some compounds show $e\phi = 2$-3 eV. This is one order of magnitude higher than required. WF values of about 1 eV were obtained in sophisticated systems like Mo-Cs and Ag-O-Cs. However, these types of electrodes have a short lifetime even in good vacuum conditions. To overcome difficulties, quantum mechanical tunneling was utilized. Tunneling through vacuum nanogap allows sufficiently large currents from the electrodes, having relatively high $e\phi$ values. It was found that image force reduces potential barrier and increases tunneling current, giving a cooling power of 100 W/cm$^2$ for $e\phi \approx 1$ eV [7] in mixed thermotunnel and thermionic regime. Electrons were filtered by collector coating, to increase the cooling coefficient [8] and conformal electrode growth technology was developed [9]. However, vacuum nanogap device appears extremely difficult to fabricate [9-11] as it requires an electrode spacing of 5-10 nm. If an $e\phi < 1$eV electrode could be obtained, poor thermionic regime can be realized at increased electrode spacing. Fabrication of wide vacuum gap, using conformal electrode technology, is much more straightforvard.

Here, we offer to reduce $e\phi$ using surface nanostructuring. The electrode is coated by the metal ridged quantum well (RQW) layer. Its operation is based on the effect of quantum state depression. Periodic ridges, fabricated on the layer surface, impose additional boundary conditions on the electron wave function. Supplementary boundary conditions forbid some quantum states for free electron, and the quantum state density in the energy $\rho(E)$ reduces. According to Pauli's Exclusion Principle, electrons rejected from the forbidden quantum states, have to occupy the states with higher $E$. As result the Fermi energy $E_F$ increases and $e\phi$ decreases [14].

The quantum state density in the RQW (Fig. 1) reduces $G$ times

$$\rho(E) = \rho_0(E)/G, \qquad (1)$$

where $\rho_0(E)$ is the density of states in a conventional quantum well layer of thickness $L$ ($a = 0$) and $G$ is the geometry factor introduced in Ref. 16. In the first approximation, for the case $a \ll L, w$ and within the range $5 < G < 10$, the following simple expression (obtained in Ref. 15) can be used

$$G \approx L/a , \qquad (2)$$

where $a$ is the ridge height and $L$ is the RQW layer thickness (Fig. 1). Density of forbidden quantum states is:

$$\rho_{FOR}(E) = \rho_0(E) - \rho_0(E)/G = \rho_0(E)(1 - G^{-1}). \qquad (3)$$

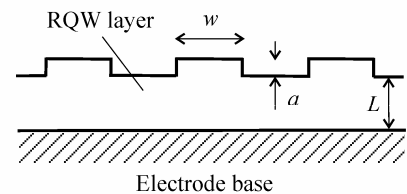

**FIG. 1**. Cross section of electrode coated by RQW

To determine the number of rejected electrons $n_{REJ}$, Eq. (3) should be integrated over the energy region in which the electrons are confined to the RQW.

$$n_{REJ} = \int_{CON} dE \rho_{FOR}(E) = (1 - G^{-1}) \int_{CON} dE \rho_0(E) = \\ = (1 - G^{-1}) n_{CON} \qquad (4)$$

Here, $n_{CON} = \int_{CON} dE \rho_0(E)$ is the number of quantum states (per unit volume) within electron confinement energy region (which depends on substrate and RQW band structures and band offset). The RQW retains quantum properties at $G$ times more width with respect to the conventional quantum well. Previously, quantum state depression was studied theoretically [16] and experimentally [17] in ridged metal films.

The objective of this work is to calculate $e\phi$ in the metal RQW layer, grown on semiconductor and metal substrates, and find out how it depends on ridge geometry and electron confinement. First, we calculate $e\phi$ values in the metal RQW forming Schottky barrier or ohmic contact with a semiconductor substrate. Next, we calculate $e\phi$ in the metal RQW grown on a metal substrate. Finally, the possibility of realization of such electrodes using conventional thermionic and other materials is discussed. Analysis was made within the limits of parabolic band, wide quantum well and degenerate electron gas approximations.

## II. WORK FUNCTION OF METAL RQW GROWN ON SEMICONDUCTOR SUBSTRATE

To maintain the uniform vacuum nanogap over the whole area, electrodes should have plane geometry and smooth surface. The simplest solution is to use semiconductor substrate as an electrode base [8] and grow a thin metal RQW layer on it.

The Energy diagram of the metal film grown on n+ type semiconductor substrate is shown in Figure 2a. We begin from the case when the difference between metal initial work function $e\phi_0$ and semiconductor electron affinity $e\chi$ is positive, i. e. $e(\phi_0 - \chi) > 0$. We presume that electron gas in metal layer is degenerate, so that all quantum states are occupied below Fermi energy $E_F$ and are empty above $E_F$. Further, suppose that the semiconductor band gap $E_g = E_C - E_V$ is wide enough, so that $e(\phi_0 - \chi) < E_g$ is satisfied. Then, electrons having energies within the region

$$\Delta E_{con}^{(0)} = E_g - e(\phi_0 - \chi) \qquad (5)$$

are confined to the metal film. Within $\Delta E_{con}^{(0)}$, quantum states for electrons are filled in the metal film and forbidden in the semiconductor substrate. Next, we fabricate ridges on the metal film surface (Fig. 2b). Additional vertical lines near the vacuum boundary in Figures. 2b and 2c depict extra boundary conditions, imposed by the ridges. Owing to quantum state depression, supplementary boundary conditions reduce the quantum state density $G$ times, within the energy region $e\chi_m - e\chi - E_g < E < e\chi_m - e\chi$, having width $\Delta E_{QSD} = E_g$. Here, we measure energies from the metal conduction band bottom $E_C^{(m)}$ and depict $e\chi_m \equiv E_F^{(0)} + e\phi_0$ as a metal conduction band width. Electrons, rejected from the forbidden quantum states within $\Delta E_{con}^{(0)}$, occupy the empty states above $E_F^{(0)}$. The Fermi level and the semiconductor band edges $E_C$ and $E_V$ move up on the energy scale. At the same time, $e\phi$ decreases and the energy region $\Delta E_{con} = E_g - e\phi + e\chi$ (Eq. (5)) gets extended. As $e\phi$ decreases, the semiconductor band edge curving follows the $e(\phi - \chi)$ value and reverses its curving direction at $e(\phi - \chi) = E_C - E_F^{(0)}$.

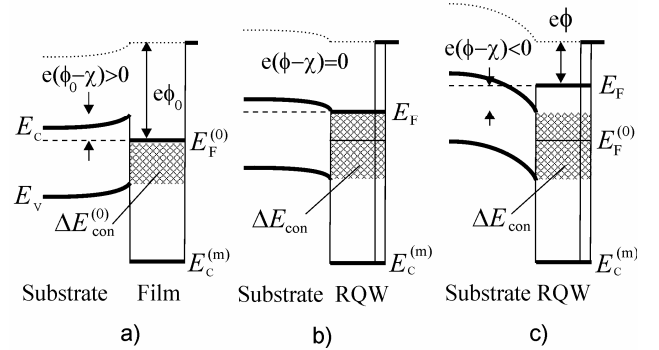

FIG. 2. a) Energy diagram of metal-semiconductor contact in the case $e(\phi_0 - \chi) > 0$, a) without periodic ridges on the surface or G=1, b) with ridges at $G = G_0$, c) with ridges at $G > G_0$. Hatch depicts confinement energy region.

The $e\phi$ value was calculated using electron number conservation in the RQW conduction band. The number of electrons rejected from the forbidden quantum states was equal to number of electrons accommodated above the initial Fermi level $n_{ACC}$. The rejected electron number $n_{REJ}$, according to Eq. (4) was

$$n_{REJ} = (1 - G^{-1}) \int_{CON} dE \rho_0(E) =$$
$$= (1 - G^{-1}) \int_{e\chi_m - e\chi - E_g}^{e\chi_m - e\varphi_0} dE \rho_0(E) \qquad (6)$$

In Eq. (6), we take into account that electrons were rejected from the energy interval $e\chi_m - e\chi - E_g < E < e\chi_m - e\phi_0$. They were accommodated in the interval $E_F^{(0)} < E < E_F$ or $e\chi_m - e\phi_0 < E < e\chi_m - e\phi$. If $n_{REJ}$ was low, the last interval fit within the $\Delta E_{QSD}$ (Fig. 2b) where the density of states is reduced Eq. (1). The accommodate electron number was equal to the number of empty quantum states between $E_F^{(0)}$ and $E_F$

$$n_{INS} = \int_{E_F^{(0)}}^{E_F} dE \rho(E) = \int_{e\chi_m - e\varphi_0}^{e\chi_m - e\varphi} dE \rho(E) =$$
$$= (1/G) \int_{e\chi_m - e\phi_0}^{e\chi_m - e\phi} dE \rho_0(E) \qquad (7)$$

Using condition $n_{REJ} = n_{ACC}$ and putting in $\rho_0(E) \propto E^{1/2}$, which is true within the limit of parabolic band approximation, we found after integration and simplification that

$$e\phi \bigg|_{e\phi \geq e\chi} = e\chi_m - e\left[ G(\chi_m - \phi_0)^{3/2} - (G-1) \times (\chi_m - \chi - E_g/e)^{3/2} \right]^{2/3} \qquad (8)$$

Here, we use 3D density of states $\rho_0(E) \propto E^{1/2}$ (wide quantum well). Further, if $n_{REJ}$ was high enough, $e(\phi - \chi)$ changed sign as shown in Figure 2c. Here, $n_{REJ}$ was calculated using Eq. (6) again. However, in this case, electrons were accommodated in quantum states from two different energy intervals. The first interval $e\chi_m - e\phi_0 < E < e\chi_m - e\chi$ was above $E_F^{(0)}$ and within the energy region $\Delta E_{QSD}$ where density of states was reduced, and the second one $e\chi_m - e\chi < E < e\chi_m - e\phi$ was above $E_F^{(0)}$ and out of $\Delta E_{QSD}$. The two intervals differ by density of states. The density was equal to Eq. (1) in the first interval and $\rho_0(E)$ in the second one. Consequently, $n_{ACC}$ was

$$\int_{E_F^{(0)}}^{E_F} dE \rho(E) = \int_{e\chi_m - e\phi_0}^{e\chi_m - e\phi} dE \rho(E) =$$
$$(1/G) \int_{e\chi_m - e\phi_0}^{e\chi_m - e\chi} dE \rho_0(E) + \int_{e\chi_m - e\chi}^{e\chi_m - e\phi} dE \rho_0(E) \qquad (9)$$

Further, equalizing Eq. (6) and Eq. (9) and repeating the above described steps we found

$$e\phi \bigg|_{e\phi < e\chi} = e\chi_m - e\left\langle \begin{array}{c} (\chi_m - \phi_0)^{3/2} + (1 - G^{-1}) \times \\ \times \left[ (\chi_m - \chi)^{3/2} - (\chi_m - \chi - E_g/e)^{3/2} \right] \end{array} \right\rangle^{2/3} \qquad (10)$$

Next, we consider the case when the difference between $e\phi_0$ and $e\chi$ was negative, i.e, $e(\phi_0 - \chi) < 0$ from the beginning (Fig. 3a). Here, $\Delta E_{con}^{(0)} = E_g$. When ridges were fabricated (Fig. 3b), the rejected electrons reduced $e\phi$. As $e\phi$ reduced $e(\phi - \chi)$ also reduced and got even more negative. Semiconductor bands curve in the direction of $E_C^{(m)}$. At the same time, $\Delta E_{con}$ width remains constant. Electrons were rejected from the interval $e\chi_m - e\chi - E_g < E < e\chi_m - e\chi$, and their number was calculated by applying this interval to Eq. (4). Electrons were accommodated in the interval $e\chi_m - \phi_0 < E < e\chi_m - e\phi$ and their number was

$$\int_{e\chi_m - e\phi_0}^{e\chi_m - e\phi} dE \rho_0(E). \quad (11)$$

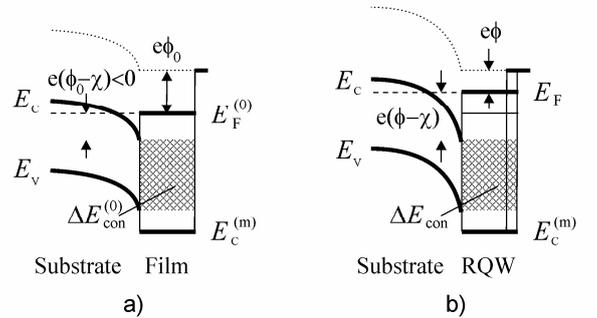

FIG. 3. Energy diagram of metal-semiconductor contact for $e(\phi_0 - \chi) < 0$, a) without periodic ridges on the surface b) with ridges. Hatch depicts confinement energy region.

Using $n_{REJ} = n_{ACC}$, and repeating the above steps, we found that Eq. (10) is true for the case $e(\phi_0 - \chi) < 0$ as well.

Finally, Eq. (8) allows the calculation of geometry factor value $G_0$, at which an ideal ohmic contact was obtained or $e(\phi - \chi) = 0$. Inserting this in Eq. (8), we found

$$G_0 = \left[(\chi_m - \chi)^{3/2} - (\chi_m - \chi - E_g/e)^{3/2}\right] \times \\ \times \left[(\chi_m - \phi_0)^{3/2} - (\chi_m - \chi - E_g/e)^{3/2}\right]^{-1}. \quad (12)$$

Analysis of metal-RQW/semiconductor contact shows that in the case $e(\phi_0 - \chi) \geq 0$, metal RQW layer $e\phi$ was calculated using Eq. (8) if $G \leq G_0$ (or $e\phi \geq e\chi$) and Eq. (10) if $G > G_0$ (or $e\phi < e\chi$). For opposite case, $e(\phi_0 - \chi) < 0$, $e\phi$ was calculated using Eq. (10) for all $G$ values (note that $G>1$ by definition).

### III. WORK FUNCTION OF METAL RQW GROWN ON METAL SUBSTRATE

In the case of metal-RQW/metal contact, electrons are confined to the material having wider conduction band (Fig. 4). Within $\Delta E_{con}^{(0)}$, quantum states for electrons are filled in the metal film and forbidden in the metal substrate (MS). There are no quantum states below MS conduction band bottom $E_C^{(S)}$ except core levels which are at $\approx -100\,\text{eV}$ and do not fall within $\Delta E_{con}^{(0)} < 10\,\text{eV}$ (values 100 eV and 10 eV are typical for metals). Let us begin from the case $e\phi_S > e\phi_0$, where $e\phi_S$ is the MS work function. Owing to WF difference, the contact potential emerges and the bottoms of conduction bands curve near the contact as shown in Figure 4a. When ridges are fabricated on the surface (Fig. 4b and 4c), some electrons are rejected from $\Delta E_{con}^{(0)}$ and accommodated above $E_F^{(0)}$. Fermi level moves up on energy scale. The $E_C^{(S)}$ follows the Fermi level (Fig. 4b). The electron confinement energy region $\Delta E_{con} = E_C^{(S)} - E_C^{(m)}$ increases. This leads to rejection of even more electrons and $e\phi$ reduction amplifies. However, with rising $E_F$, number of states $\int dE \rho_0(E) \propto E^{3/2}$ above $E_F$ increase more rapidly than $n_{REJ}$ (as $E_F > E_C^{(S)}$) and at some $e\phi$ value, equilibrium is maintained.

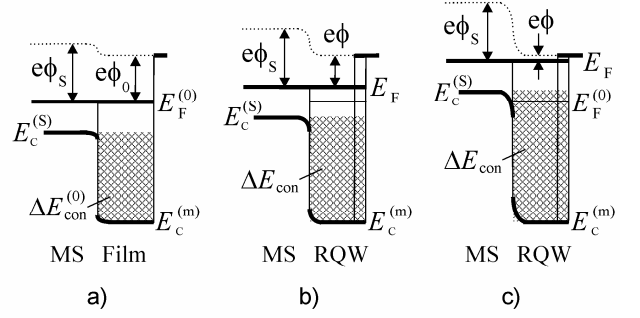

**FIG. 4.** Energy diagram of metal-metal contact for the case $e\phi_S > e\phi_0$. a) Without periodic ridges, b), c) with ridges. MS depicts metal substrate. Hatch depicts confinement energy region.

Let us consider the case of low $n_{REJ}$ or $E_C^{(S)} < E_F^{(0)}$ first (Fig. 4b). To find $e\phi$, we use electron number conservation in RQW conduction band again. Electrons were rejected from the energy region $0 < E < E_C^{(S)}$ or $0 < E < e\chi_m - e\phi + e\phi_S - e\chi_S$ where, $e\chi_S \equiv E_F^{(S)} + e\phi_S$ was the width of the substrate conduction band ($E_F^{(S)} = E_F - E_C^{(S)}$ was substrate Fermi energy). The $n_{REJ}$ was calculated by applying the last region to Eq. (4). Electrons were accommodated in the energy interval $E_F^{(0)} < E < E_F$ or $e\chi_m - e\phi < E < e\chi_m - e\phi_0$, where density of state was $\rho_0(E)$. The $n_{ACC}$ was calculated via Eq. (11). Using $n_{REJ} = n_{ACC}$ and integrating, we obtain the following equation for $e\phi$

$$(1 - G^{-1})(e\chi_m - e\phi + e\phi_S - e\chi_S)^{3/2} = \\ = (e\chi_m - e\phi)^{3/2} - (e\chi_m - e\phi_0)^{3/2}. \quad (13)$$

In the case of high $n_{REJ}$ or $E_C^{(S)} > E_F^{(0)}$ (Fig. 4c), electrons were rejected from the interval $0 < E < e\chi_m - e\phi_0$ and we applied this interval to Eq. (4). They were accommodated in two intervals having diverse density of states. The first interval $e\chi_m - e\phi_0 < E < e\chi_m - e\phi + e\phi_S - e\chi_S$ was above $E_F^{(0)}$ and within the $\Delta E_{con}$, and the second one $e\chi_m - e\phi + e\phi_S - e\chi_S < E < e\chi_m - e\phi$ was above $E_F^{(0)}$ and out of $\Delta E_{con}$. The two intervals differ in the density of states. Density was equal to Eq. (1) in the first interval and $\rho_0(E)$ in the second one. The accommodated electron number was

$$\int_{E_F^{(0)}}^{E_F} dE \rho(E) = \int_{e\chi_m - e\phi_0}^{e\chi_m - e\phi} dE \rho(E) =$$
$$= (1/G) \int_{e\chi_m - e\phi_0}^{e\chi_m - e\phi + e\phi_S - e\chi_S} dE \rho_0(E) + \quad (14)$$
$$+ \int_{e\chi_m - e\phi + e\phi_S - e\chi_S}^{e\chi_m - e\phi} dE \rho_0(E)$$

Using $n_{REJ} = n_{ACC}$ and, we obtain Eq. (13) again for the last case.

Analysis of Eq. (13) shows that in some cases $e\phi$ can be reduced to zero and even acquire negative values, hypothetically. However, negative $e\phi$ can not be realized since, in that case, at least one electron leaves RQW to vacuum. Simultaneously, the electrode as a whole charges positively, owing to electrical neutrality. This leads to shifting down of band bottoms $E_C^{(m)}$ and $E_C^{(S)}$ on the energy scale. At the same time, their difference $\Delta E_{con} = E_C^{(S)} - E_C^{(m)}$ does not increase and no additional electron is rejected. Naturally, vacuum level retains its position, which implies that both the conduction bands widen a little bit as their bottoms shift down. In the wider band, a new quantum state emerges at the band top. The electron that left return (following image force) and occupy the emerged quantum state. As result, $e\phi$ increases back to zero value. Some details on this mechanism can be found in Ref. 16. Related mechanisms were studied in negative electron affinity semiconductors[18].

The energy diagram for the opposite case $e\phi_S < e\phi_0$ is shown in Figure 5a. Initially, the vacuum level curves in the opposite direction, the contact potential has opposite sign and the conduction band bottoms curve in opposite directions. In this case, a small reduction in $e\phi$ results in the flattening of the conduction band bottoms. The width of $\Delta E_{con}$ reduces instead of increasing. If initial $n_{REJ}$ (rejected from $\Delta E_{con}^{(0)}$) was low, the system remains in the state $e\phi_S - e\phi_0 < 0$. However, if initial $n_{REJ}$ was high enough to reduce $e\phi$ below $e\phi_S$, the band bottom curving direction changes (Fig. 5b). Energy diagram becomes analogous to the one shown in Figure 4a and $e\phi$ can be calculated using Eq. (13) again. Further, we seek the initial $n_{REJ}$, required to change the sign of $\phi_S - \phi$. The following approximation was used to determine this number. It was assumed that the charge depletion depth has the same values in the two metals and is much less than the thickness of the RQW layer (electron concentration is thought to be equal in two metals). Within this approximation the electric field distribution around the contact area was symmetric and electron confinement energy region was $E_C^{(m)} < E < E_C^{(S)} + e\phi^{(-)}$, where $\phi^{(-)} \approx (\phi_0 - \phi_S)/2$. Here, we did not take into account the quantum states inside the triangular quantum well below $E_C^{(m)}$, assuming that the triangular well width was low with respect to RQW. Electrons, rejected from the above interval, should fill the interval $F_F^{(0)} < E < F_F^{(0)} + e(\phi_0 - \phi_S)$ to equalize WFs and flatten bang bottoms. According to Eq. (4)

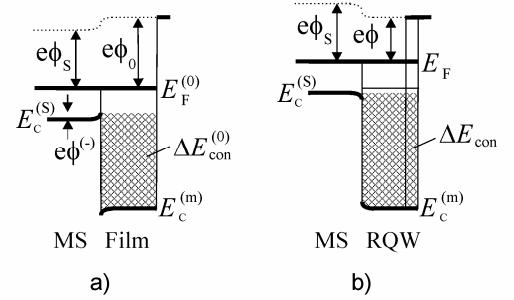

**FIG. 5**. Energy diagram of metal-metal contact for the case $e\phi_S < e\phi_0$. a) Without periodic ridges, b), c) with ridges. MS depicts metal substrate. Hatch depicts confinement energy region.

$$(1 - G^{-1}) \int_{E_C^{(m)}}^{E_C^{(S)} + e\phi^{(-)}} dE \rho_0(E) = \int_{E_F^{(0)}}^{E_F^{(0)} + e(\phi_0 - \phi_S)} dE \rho_0(E). \quad (15)$$

Integrating Eq. (15) and simplifying, we found threshold geometry factor

$$G_{th} = (\chi_m - \chi_S + \phi_S/2 - \phi_0/2)^{3/2} \times$$
$$\times \left[ (\chi_m - \chi_S + \phi_S/2 - \phi_0/2)^{3/2} + (\chi_m - \phi_0)^{3/2} - (\chi_m - \phi_S)^{3/2} \right]^{-1}. \quad (16)$$

If $G \geq G_{th}$ system ends with reduced $e\phi$ in the RQW, and if $G < G_{th}$ system remains in the initial state $e(\phi_S - \phi_0) < 0$.

Analysis of metal-RQW/metal contact show that if initially $e\phi_S > e\phi_0$, $e\phi$ of RQW layer was reduced and its value was calculated using Eq. (13). If initially opposite relation $e\phi_S < e\phi_0$ was realized, the last was true only in the case $G \geq G_{th}$. Other wise the reduction was irrelevant.

## IV. MATERIAL PAIRS FOR RQW COATED ELECTRODE

The thickness of a conventional metal quantum well is only 1-5 nm [19]. Periodic ridge fabrication on such thin layer surface seems complicated. However, owing to reduced quantum state density, metal RQW layer retains quantum properties at $G$ times more thickness (as found in Ref. 15) with respect to a conventional metal quantum well. This makes ridge fabrication straightforward. In some cases, layer thickness can be increased even more than $G$ times. As Figures 3, 4 and 5 show, confined electrons have energies $E < E_F^{(0)} - K_B T$ (here, $K_B$ is Boltzmann constant and $K_B T \approx 25$ meV $<< E_F^{(0)}$ for $T$=300K). For these energies, Fermi-Dirac distribution function $f(E, E_F, T) \approx 1$. Consequently, electron scattering on phonons and lattice defects is limited. Such scattering requires the exchange of a small portion of energy with environment, which is quantum mechanically restricted as all nearby quantum states are occupied. The mean free path of such electrons is very large and is limited only by structural defects, such as grain boundaries. In epiatxial films, such electrons allow phase coherence at large distances (in transverse direction) and metal RQW thickness can be increased.

Regular semiconductor materials can be used as base substrate. Thermionic and thermotunnel converters are high current low voltage devices. Consequently, the main limitation is electrical conductivity. Equations (8) and (10) show that wide band gap material allows more reduction in $e\phi$. Unfortunately, it is problematic to achieve low resistivity in such materials. The most promising seems GaN in which relatively low resistivity was obtained. Other possible substrate materials are GaAs and Si. Table I shows possible donor concentrations $N_D$ for GaN [20], Si, GaAs [21] and corresponding electrical resistivity $r$ and heat conductivity $\kappa$. Metals, Ni and Mo, are also included since they are frequently used in thermionic converters. Table I also shows the dissipated power per cm$^2$ area $P_d$, and temperature gradient $\nabla T$ over 1 mm thick substrate, calculated for typical current density (10 A/cm$^2$) and heat flux (10 W/cm$^2$). Power dissipation in GaN substrate is considerable (note that $\nabla T$ is low for GaN). On the other hand, GaN has a wide band gap and allows lower $e\phi$ values. Possible solution is to grow thin GaN epitaxial layer on the GaAs or Si substrates. Such bi-layer substrate ensures low power loss together with low $e\phi$ in the metal RQW layer. Fortunately, GaN can be grown epitaxially on both GaAs and Si substrates [22].

**TABLE I.** Parameters of electrode base materials. The $P_d$ and $\nabla T$ are given for 1 mm thick substrate.

| Substrate material | $E_g$ (eV) | $e\chi$ (eV) | $N_D$ (cm$^{-3}$) | $r$ ($\Omega$ cm) | $P_d$ (mW) | $\kappa$ (Wcm$^{-1}$K$^{-1}$) | $\nabla T$ (K) |
|---|---|---|---|---|---|---|---|
| Si | 1.12 | 4.05 | 8 10$^{20}$ | 2 10$^{-4}$ | 2 | 1.6 | 0.6 |
| GaAs | 1.42 | 4.07 | 8 10$^{19}$ | 1 10$^{-4}$ | 1 | 0.5 | 2.0 |
| GaN | 3.20 | 4.10 | 1 10$^{19}$ | 7 10$^{-3}$ | 70 | 5.0 | 0.2 |
| Mo | | | | 5.3 10$^{-6}$ | 5.3 10$^{-2}$ | 1.4 | 0.7 |
| Ni | | | | 6.2 10$^{-6}$ | 6.2 10$^{-2}$ | 0.9 | 1.1 |

Most technological metals have $e\phi_0 > 4$ eV and form Shottky barriers with above semiconductor materials i.e. $e(\phi_0 - \chi) > 0$. Table II shows parameters of some metals frequently used as electrode materials (other interesting metals are included as well), collected from the literature: $e\phi_0$ from Ref. 23 and $E_F^{(0)}$ from Ref. 24. It also contains data on $E_F^{Band}$, which is the Fermi energy obtained from *ab initio* calculations and experiments [25-30]. Parameter $E_F^{Band}$ differs more or less from $E_F^{(0)}$. We used $E_F^{Band}$ instead of $E_F^{(0)}$ for Mo, W, Ni and Pt since no $E_F^{(0)}$ data was found. This substitution is acceptable as far as a parabolic band is good approximation for these metals. The $e\phi$ values presented in the table were calculated using Eq. (10). WF values were reduced by $\approx 1$ eV on Si, by $\approx 1.2$ eV on GaAs and by $\approx 2.3$ eV on GaN substrates. Reduction is limited by the $\Delta E_{con}$ scale. The last can only be increased by increasing band gap. However, wide band gap substrates have high resistivity and were not acceptable for thermotunnel and thermionic devices. Obviously 1-2.3 eV reduction in $e\phi$ was not enough for cold emission, but it was still interesting since the plain Mo and Ni electrodes, coated with ultra thin layer of Cs atoms, show very low WF (1-1.5 eV). WF reduction by less than mono layer of Cs atoms is a surface effect.

**TABLE II.** Characteristic energies for some metals and values of $e\phi$, calculated for $G=10$

| RQW Mat. | $e\phi_0$ (eV) | $E_F^{(0)}$ (eV) | $E_F^{Band}$ (eV) | $e\phi$ (eV) on Si | GaAs | GaN |
|---|---|---|---|---|---|---|
| Ag | 4.26 | 5.48 | 7.5[a] | 3.32 | 3.10 | 2.02 |
| Nb | 4.30 | 5.32 | 5.5[b] | 3.36 | 3.14 | 2.06 |
| W | 4.55 |  | 6.7[c] | 3.58 | 3.34 | 2.14 |
| Cu | 4.65 | 7.0 | 9.1[a] | 3.65 | 3.43 | 2.20 |
| Mo | 4.60 |  | 5.0[d] | 3.63 | 3.41 | 2.32 |
| Au | 5.10 | 5.53 | 9.4[a] | 4.09 | 3.85 | 2.64 |
| Ni | 5.15 |  | 5.0[e] | 4.14 | 3.90 | 2.81 |
| Pt | 5.65 |  | 10[f] |  |  | 2.95 |

[a]Ref. 27, [b]Ref. 28, [c]Ref. 26, [d]Ref. 25, [e]Ref. 29, [f]Ref.30.

At the same time, quantum state depression is not a surface effect. Most probably, the two mechanisms of WF reduction will sum up and result in an $e\phi$ considerably less than 1 eV. The same is true for Ag-Ba and W-Li electrodes.

Borides have $e\phi_0 < 4$ eV and for them $e(\phi_0 - \chi) < 0$ (Fig. 3). The most frequently used is Lanthanum hexaboride LaB$_6$, which shows $e\phi$=2-3.2 eV. Fermi energy for LaB$_6$ is $E_F^{Band}$=10 eV [13]. Inserting these values in Eq. (10) gave $e\phi$ = 0-0.85 eV for ridged LaB$_6$ layer on GaN substrate, $e\phi$ = 0.94-2.05 eV on GaAs substrate and $e\phi$ = 1.15-2.28 eV on Si substrate ($G=10$ was used in all cases). These values were low enough for thermotunnel and thermionic devices operating at room temperatures.

Geometry factors, at which ideal ohmic contact was obtained, were calculated using Eq. (12). They were in the range $G_0 = 1-5$ for most material pairs. Such $G_0$ can easily be obtained in practice. Exceptions were Ni/Si ($G_0 = 59$) and Au/Si ($G_0 = 16$) pairs.

Next, we consider metal substrates (Fig. 4 and 5). Electron confinement energy region emerges only if $E_F^{Band}$ of a substrate material is less than that of a RQW material. Nickel and Molybdenum were good choices for substrates as they have low $E_F^{Band}$. Au, Pt, Cu materials were suitable for RQW. They have high $E_F^{Band}$ and at the same time can be grown epitaxially on Ni substrate [32-34]. The $e\phi$ in RQW depends on material parameters and $G$ according to Eq. (13). To determine the geometry factor, needed for $e\phi$ = 0.5 eV in RQW, material parameters from Table II together with $e\phi$ = 0.5 eV were put in Eq. (13). Results were $G_{e\phi=0.5}$ =8.2 for the case of Cu/Ni, $G_{e\phi=0.5}$ =7.8 for Au/Ni and $G_{e\phi=0.5}$ =6.5 for Pt/Ni. These values were low enough to be realized in practice. One more interesting RQW material is TiN since it has high $E_F^{Band}$ (in fact it has two bands) [35]. However, lattice mismatch introduces problems in TiN epitaxial growth.

Threshold geometry factor $G_{th}$ for Ni and Mo substrates was calculated using Eq. (16). We got $G_{th}$ =1.1 for Pt/Ni, $G_{th}$ =1.3 for Au/Mo and $G_{th}$ =1.6 for Pt/Mo. For other pairs, $G_{th}$ = 1.

Analysis, made on the basis of material parameters, shows that low $e\phi$ electrodes can be obtained using RQW layer made from conventional thermionic materials. Both semiconductor and metal substrates can be used to obtain low $e\phi$. However, metal substrate is preferable as it allows low $e\phi$ for a broad range of materials. Dissipated power and temperature gradient calculations show that both semiconductor and metal substrates, coated with metal RQW layer, can be used as electrodes for thermionic and thermotunnel converters.

## V. CONCLUSIONS

Low work function was obtained in the metal ridged quantum well having reduced quantum state density. Metal RQW layers, grown on semiconductor and metal substrates, were analyzed. Electron confinement to the RQW was essential in both cases. When using semiconductor substrate, wide band gap material allows more electron confinement and lower values of resulting $e\phi$. Dependence of $e\phi$ on the band gap was analyzed for a number of cases and the corresponding formulae derived. When using metal substrate, materials with low Fermi energy allow more electron confinement and lower values of resulting $e\phi$. If initial $e\phi_0$ in the metal layer was less than in the substrate, $e\phi_0 < e\phi_S$, considerable $e\phi$ reduction was obtained for all metal pairs, ensuring electron confinement. In the opposite case, $e\phi_0 > e\phi_S$, low $e\phi$ was obtained only if the geometry factor exceeded some threshold value.

**ACKNOWLEDGMENTS**
The author thanks L. Tsakadze for useful discussions and the Physics Department of New York University, where part of this work was done, for their hospitality.